\documentclass[12pt]{iopart}
\def\lsim{\mathrel{\rlap{\lower4pt\hbox{\hskip1pt$\sim$}}
    \raise1pt\hbox{$<$}}}		 
\begin{document}
\jl{4}
\title[Radiative tail from the quasielastic peak in DIS
of polarized leptons off polarized $^3$He]
{Radiative tail from the quasielastic peak in deep inelastic
scattering of polarized leptons off polarized $^3$He}
\author {I.Akushevich$^{1,2}$, D.Ryckbosch$^3$, N.Shumeiko$^1$,
A.Tolkachev$^1$}
\address{ $^1$ National Center of Particle and High Energy Physics,
 Minsk Belarus}
\address{ $^2$ NC Central University NC and TJNAF VA USA}
\address{ $^3$ Gent University, Belgium}
\begin{abstract}
The contribution of the radiative tail from the quasielastic peak
to low order radiative correction to deep inelastic
scattering of
polarized leptons by polarized $^3$He was calculated within the sum
rules formalism and $y$-scaling hypothesis. Numerical analysis
was carried out under the conditions of HERMES experiment.
\end {abstract}
\submitted
\maketitle

\input{epsf}

\section {Introduction }

There is great interest in the
investigation of the spin structure of	nucleons
in modern high energy physics.
Experiments
on deep inelastic scattering (DIS) of polarized leptons by
polarized light nuclei are among the most powerful sources to
obtain such information. The interpretation of results
of these experiments requires the correct treatment of radiative
corrections (RC), to minimize their contribution to the systematic errors.

It is well known that there are three channels for scattering of a
virtual photon by a nucleus, depending on the energy transfer
$\nu=E_1-E_2$, with $E_{1}(E_{2})$ the initial	(scattered)
lepton	energy: elastic, quasielastic and inelastic. Both
$\nu$ and
the momentum transfer (squared) $Q^{2}=-q^{2}$	are fixed by
kinematical  conditions  on  the  Born	level and, consequently,   also the
scattering channel is  fixed.	However, on the  RC level the
uncertainty in the energy of  the radiated photon makes $Q^2$  and
$\nu$  indefinite  and	as  a  result  each  of the three channels
contributes to the cross  section. Whereas  contributions
from the
elastic and  inelastic tails  are well	studied, the  contribution
from the quasielastic  tail was  not thoroughly  investigated.	 Up to
now there are only approximate results, for a few target nuclei,
and based  on phenomenological models\cite{ST,BA}  using the  Mo and
Tsai \cite{MT} formalism. For $^3$He no results were published so far.
Only the general framework exists
for the case of polarized reactions \cite{ASh}. In this paper
the cross section and
the polarization asymmetry  of DIS of polarized
charged leptons from polarized targets (H, D, $^3$He)  have been
investigated both on the Born
level and taking into account RC.
The purpose of the present paper is to develop the methods that are necessary
to correctly treat the radiative tail from the quasielastic peak.
This is then applied to the specific case of a $^3$He target,
using approximate (due to the lack of
experimental data), but model independent approaches.
Numerical results are presented relevant for the kinematics
of the HERMES experiment.

The paper is organized as follows. In section 2 we derive the basic
formulae and consider $y$-scaling. Section 3 is devoted to the
implementation of sum rules approach to the calculation of RC.
Finally, in section 4 we present the numerical analysis.

\section{Radiative tail from the quasielastic peak}

The model independent RC of the lowest order can be written as
a sum of brems\-strahlung and loop effects:
\begin {equation}
\sigma	= \sigma _{in}+ \sigma _{el}+ \sigma _{q}+ \sigma _{v}.
\label{eq1}
\end {equation}
Each  $\sigma  $  denotes a double  differential  cross  section
$d^2\sigma / dxdy$, with $x,y$ the  usual scaling
variables;
$\sigma  _{in,el,q}$  are  the	contributions  of the radiative
inelastic, elastic and quasielastic (QRT) tails.
The term $\sigma_v$
contains the contribution of virtual photon radiation and
vacuum polarization effects.

The exact contribution of the QRT
to total RC of the lowest order
in the case of DIS of polarized leptons off polarized
nuclei is given by
\begin{equation}
 \sigma_q =- {\alpha ^{3}S_{x}S\over \pi \lambda
_{s}}\int {d^3k\over k_0}
{1\over (q-k)^4}
{\cal L}_{\mu \nu}W_{\mu \nu}{(p,q-k)}.
\label {r1}
\end{equation}
Invariants are defined as usual:
\begin{equation}
\begin{array}{c}
\displaystyle
S = 2k_{1}p,\; X = 2k_{2}p = (1-y)S,\; Q^2 = -(k_{1}-k_{2})^{2} ={ xyS},
\\[0.5cm]\displaystyle
S_{x}= S-X,\; \lambda _{s}=
S^{2}-4m^{2}M^{2},
\end{array}
\end{equation}
where  $k_{1}(k_{2}),m$  are the initial (final) lepton momentum
and its  mass respectively, $k$ is the momentum
of the radiated photon and $p,M$ are momentum and the mass of the
initial nucleon.

The leptonic tensor
we use in (\ref{r1}) is standard for bremsstrahlung processes and
includes spin averaged and spin dependent parts \cite{Ak}.
For the hadronic tensor we have
\begin {eqnarray}\label{ht}
W_{\mu \nu }(p,q)&=& - (g_{\mu \nu } + {q_{\mu }q_{\nu }\over Q^{2}})
{\cal F} _{1}
+\frac{1}{M^2}
{(p_{\mu }+ {pq \over Q^{2}} q_{\mu })(p_{\nu }+ {pq \over
Q^{2}}
q_{\nu })}{\cal F} _{2}+
\nonumber\\&&
+ \frac{i}{M}{
 \epsilon _{\mu \nu \alpha \beta }q_{\alpha }\eta _{\beta }}{\cal F} _{3}
-\frac{i}{M^3}{
 \epsilon _{\mu \nu \alpha \beta }q_{\alpha }p _{\beta }
 (q\eta) }{\cal F} _{4},
\end {eqnarray}
where $\eta$ is a target polarization vector.
The quantities ${\cal F} _{i}$	are  defined  as  some	combinations  of the
nucleon structure functions.
The dependence of the hadronic tensor on $pq$  and the	polarization
of the beam ($P_L$) and target ($P_N$)
is also included in the ${\cal F} _{i}$. Defining $\epsilon =M^{2}/pq$	we
 have
for the various ${\cal F} _{i}$:
\begin {equation}
\begin {array}{l}
{\cal F} _{1}= F_{1} , \ \ \ \ \ {\cal F} _{3}= P_{N}\;\epsilon
\pmatrix{g_{1}+g_{2}}, \\[0.5cm]
{\cal F} _{2}= \epsilon F_{2} , \ \ \
{\cal F} _{4}= P_{N}\;\epsilon ^{2}\;\;g_{2}.
\end {array}
\end {equation}
Definitions of
SF $F_{1,2}$ and $g_{1,2}$ are the same as in ref. \cite {HJM}.

As was shown in \cite{ASh} formula (\ref{r1}) leads to the
following expression:

\begin {equation}
\sigma_{q}=
- {\alpha ^{3}y\over A}\int \limits^{\tau_{max}}_{\tau_{min}}
d\tau\sum^{4}_{i=1}
 \sum^{k_{i}}_{j=1} \theta _{ij}(\tau)
 \int  dR
 {R^{j-2}\over {\tilde Q}^2}{\cal F}_{i}^{q}(R,\tau).
\label {in}
\end {equation}
The variables $R$ and $\tau$ are defined as
\begin {equation}
\begin {array}{c}
\displaystyle
R = 2pk, \qquad \tau = {({\tilde Q}^2 - Q^2)/R},
\\[0.5cm]
 \displaystyle
\qquad {\tilde Q}^2=-(k_{1}-k -k_{2})^{2}=Q^2 + R\tau,
\\[0.5cm]
 \displaystyle
\tau _{max,min} = {S_{x} \pm \sqrt {S^{2}_{x}+ 4M^{2}Q^2}\over
2M^{2}}
\end {array}
\label{exact}
\end {equation}
and $k_i=(3,3,4,5)$.
The
explicit form of the functions
$\theta _{ij}(\tau)$
can be found in Appendix B of ref. \cite{ASh}.

The quantities ${\cal F} ^q_i$ could be obtained in terms of quasielastic
structure functions (so-called response functions, see Appendix A
of \cite{AT}
for explicit results), which are peaked around
$\nu_q
=Q^2/2M$. Due to the absence of sufficient experimental data this
fact is normally used as the basis of the peaking approximation:
the response functions
are estimated at the position of the
peak, and a
subsequent integration over the peak leads to results in terms
of suppression
factors $S_{E,M,EM}$ or (see discussion below) of sum rules for electron-nucleus scattering
\cite{LLS}:
\begin {equation}
\sigma^{q}_1=
- {\alpha ^{3}y\over A}\int\limits^{\tau_{max}}_{\tau_{min}}d\tau
 \sum^{4}_{i=1}\sum^{k_{i}}_{j=1} \theta _{ij}(\tau){
 2M^{2} R^{j-2}_{q}\over (1+\tau)(Q^2+R_{q}\tau)^{2}}{\cal F}
^{q}_{i}(R_{q},\tau),
\label{eq23q}
\end {equation}
where $R_{q}=(S_x - Q^2) / (1+\tau)$.

In particular, for a $^3$He target one finds:
\begin {equation}
\begin {array}{l}
\displaystyle
{\cal F} ^{q}_{1}= \eta (\mu_n^2+2\mu_p^2)S^u_M,\\[0.5cm]
\displaystyle
{\cal F} ^{q}_{2}=\displaystyle{\eta
(\mu_n^2+2\mu_p^2)S^u_M+(e_n^2+2e_p^2)S_E \over
1+\eta },\\[0.5cm]
\displaystyle
{\cal F} ^{q}_{3}=
\displaystyle{P_N}2(P_ne_n\mu_n+2P_pe_p\mu_p)S_{EM},
\\[0.5cm]
\displaystyle
{\cal F} ^{q}_{4}=  {P_N\over 4}{(P_ne_n\mu_n+2P_pe_p\mu_p)S_{EM}-
(P_n\mu_n^2+2P_p\mu_p^2)S^p_M \over 1+\eta }
.
\end {array}
\label {qf}
\end {equation}
$P_p$ and $P_n$ are the effective proton and neutron polarization in $^3$He
and $e, \mu_{p,n}$ are the electric and magnetic formfactors for proton and
neutron, $\eta ={\tilde Q}^2/4M^{2}$.

Explicit forms of the suppression factors (or functions ${\cal F}^q_i$)
depend on a nuclear model for quasielastic scattering. Assuming the
validity of $y$-scaling for quasielastic scattering \cite{West}
we have \cite{Tho}
\begin {equation}
S^u_M=S^p_M=S_E=S_{EM}=F(\nu_q),
\label {0004}
\end {equation}
where $F(\nu_q)$ is a scaling function evaluated at the
quasielastic peak.
Neither a fit to experimental data nor a simple model for
scaling function exist today. In the current version of the radiative
correction code POLRAD, an
extrapolation of a Fermi gas model \cite{Wal,Mon} is used which
is really only applicable in heavier nuclei.

\section{Quasielastic radiative tail and sum rules}

Another possibility to obtain explicit forms for the suppression
factors is to
use the sum rules for electron-nucleus scattering \cite{LLS,OrTr}.

\begin {equation}
m _{I}^{n}=
\int d\nu (\nu-\nu_q)^n
R^{I},
\label{sumrules}
\end {equation}
where $R^I$ ($I={L,T,T',TL'}$) are quasielastic response functions
\cite{LLS}.

The functions ${\cal F}_i$ are linear combinations of these quasielastic
response functions, which are supposed to have a form of a  peak over
transfer energy $\nu$. This fact can be used for construction of some
general expansion of coefficients in front of $R^{L,T,T',TL'}$ over
$(\nu-\nu_q)$. Integration over $dR=2Md\nu$ gives us just sum rules
(\ref{sumrules})

\begin{eqnarray}\label{razlo}
 \int  dR
 {R^{j-2}\over {\tilde Q}^2}{\cal F}_{i}^{q}(R,\tau)
&=&\int  dR {R^{j-2}\over {\tilde Q}^2} \sum_I C_i^I R^I
=\\&=&\sum_{n,I} C^I_{in} \int  d\nu (\nu-\nu_q)^n  R^I
=\sum_{n,I} C^I_{in} m_I^n
\nonumber
\end{eqnarray}

The sum rules are defined as a sum over final
states of the cross
section for the inclusive elastic and quasielastic scattering. Using the
completeness property of final state nuclear wavefunctions the
sum rules for electron-nucleus scattering are obtained as a sum of
contributions of
elastic and quasielastic scattering integrated over energy.

Since inclusive scattering is defined by two kinematical variables,
sum rules correspond to the choice of external variables being some function 
of $Q^2$ and $\nu$. In our case we
have to keep $\tau$ as an external variable. However if to restrict the
consideration to the lowest order of expansion (\ref{razlo})
the difference between the sum rules due to the choice of the external 
variable is of the next neglected order, as it is always possible 
to recalculate it using $\nu$ taken at the peak.

For the calculation we use sum rules obtained in the paper \cite{LLS}, but subtracting 
an elastic contribution.
The
cross section for quasielastic
scattering needed here is then
obtained in terms of well defined model independent quantities (sum
rules) and experimentally measured quantities (elastic formfactors):
\begin {equation}
{d\sigma_{q}\over d\Omega}
={d\sigma_{SR}\over d\Omega}
-{d\sigma_{el}\over d\Omega},
\label {0202}
\end {equation}
where $\Omega$ is lepton solid angle.

Using the explicit results for sum
rules
given in e.g. \cite{LLS} and performing explicit subtraction of elastic
contribution (\ref{0202})
 the suppression factors can be obtained in the following form:
\begin {equation}
\begin {array}{l}
\displaystyle
S_M^{u}=1-{2\mu_p^2T_q \over 2\mu_p^2+\mu_n^2}-
{v_{TA} \over v_T}{\eta_A \over \eta}{f \over f_A}
{4G_M^2 \over 2\mu_p^2+\mu_n^2},
\\[0.5cm]\displaystyle
S_E=1+{2(e_p^2+2e_pe_n)T_q \over 2e_p^2+e_n^2}
-{\rho_A \over \rho}{f \over f_A}
{4G_E^2 \over 2e_p^2+e_n^2},
\\[0.5cm]\displaystyle
S_M^{p}=1
-{v'_{TA} \over v'_T}{\eta_A \over \eta}{f \over f_A}
{4G_M^2 \over \mu_p^2P_p+\mu_n^2P_n},
\\[0.5cm]\displaystyle
S_{EM}=1+{2e_p\mu_nT_q \over e_p\mu_pP_p+e_n\mu_nP_n}
{Q'^2_A \over Q'^2}{M \over M_A}{q \over q_A}{f \over f_A}
{4G_EG_M \over \mu_pe_pP_p+\mu_ne_nP_n},
\end {array}
\label {0005}
\end {equation}
where we use the structure function $T_q$, which is the Fourier
transform of the two-body density matrix
(ref.\cite{LLS}) and take for $T_q$
results of ref.\cite{Shia}.
The last
terms in righthandside of Eqs.(\ref{0005}) correspond to elastic
scattering off
$^3$He and are obtained in terms of nuclear formfactors
$G_E$ and
$G_M$. The other
stem from the sum rule
expressions (14a-d) of ref.\cite{LLS} and depend
on the nucleon formfactors $e, \mu_{p,n}$. Kinematical quantities appearing
in the last terms ($M_A$ is the nucleus mass) are defined as

\begin{figure}[t]
\hspace{2.5cm}
\begin{picture}(160,160)
\put(10,120){\makebox(0,0){$F(\nu_q)$}}
\put(210,120){\makebox(0,0){$S_E$}}
\put(10,-55){\makebox(0,0){$S^u_M$}}
\put(220,-55){\makebox(0,0){$S^p_M
$}}
\put(140,-210){\makebox(0,0){$\tilde Q^2$}}
\put(340,-210){\makebox(0,0){$\tilde Q^2$}}
\put(140,-35){\makebox(0,0){$\tilde Q^2$}}
\put(340,-35){\makebox(0,0){$\tilde Q^2$}}
\put(150,120){\makebox(0,0){$a)$}}
\put(350,120){\makebox(0,0){$b)$}}
\put(150,-55){\makebox(0,0){$c)$}}
\put(350,-55){\makebox(0,0){$d)$}}
\put(0,-320){
\epsfxsize=13cm\epsfysize=18cm\epsffile{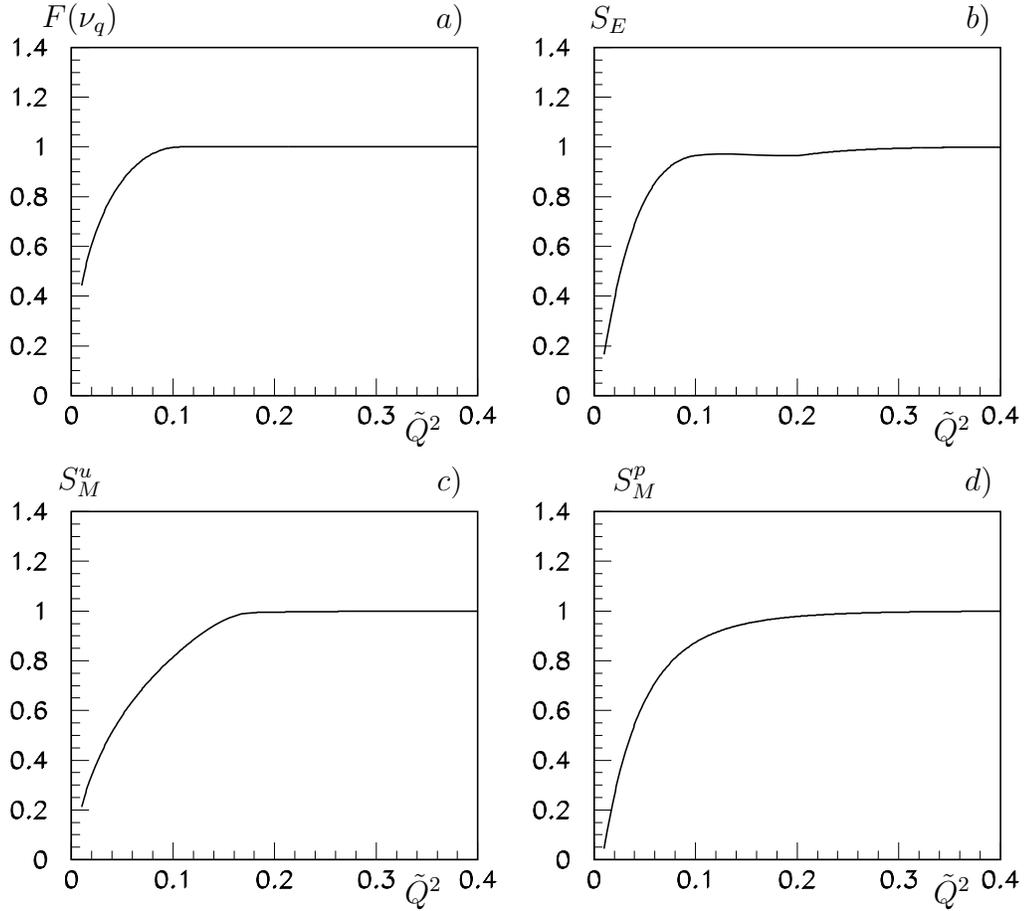}
}
\end{picture}
\vspace{7.5cm}
\caption{
The ${\tilde Q}^2$ dependence of suppression
factors calculated within the y-scaling hypothesis (1a) and
within the sum
rules formalism (1b - 1d).
}
\label{figsup}
\end{figure}

\begin{figure}[t]
\hspace{2.5cm}
\begin{picture}(160,160)
\put(10,120){\makebox(0,0){$\delta_y (\%)$}}
\put(210,120){\makebox(0,0){$\delta_{sr} (\%)$}}
\put(30,-55){\makebox(0,0){
$\delta_y - \delta_{sr},
(\%)$
}}
\put(150,-210){\makebox(0,0){$x$}}
\put(350,-35){\makebox(0,0){$x$}}
\put(150,-35){\makebox(0,0){$x$}}
\put(150,120){\makebox(0,0){$a)$}}
\put(350,120){\makebox(0,0){$b)$}}
\put(150,-55){\makebox(0,0){$c)$}}
\put(0,-320){
\epsfxsize=13cm\epsfysize=18cm\epsffile{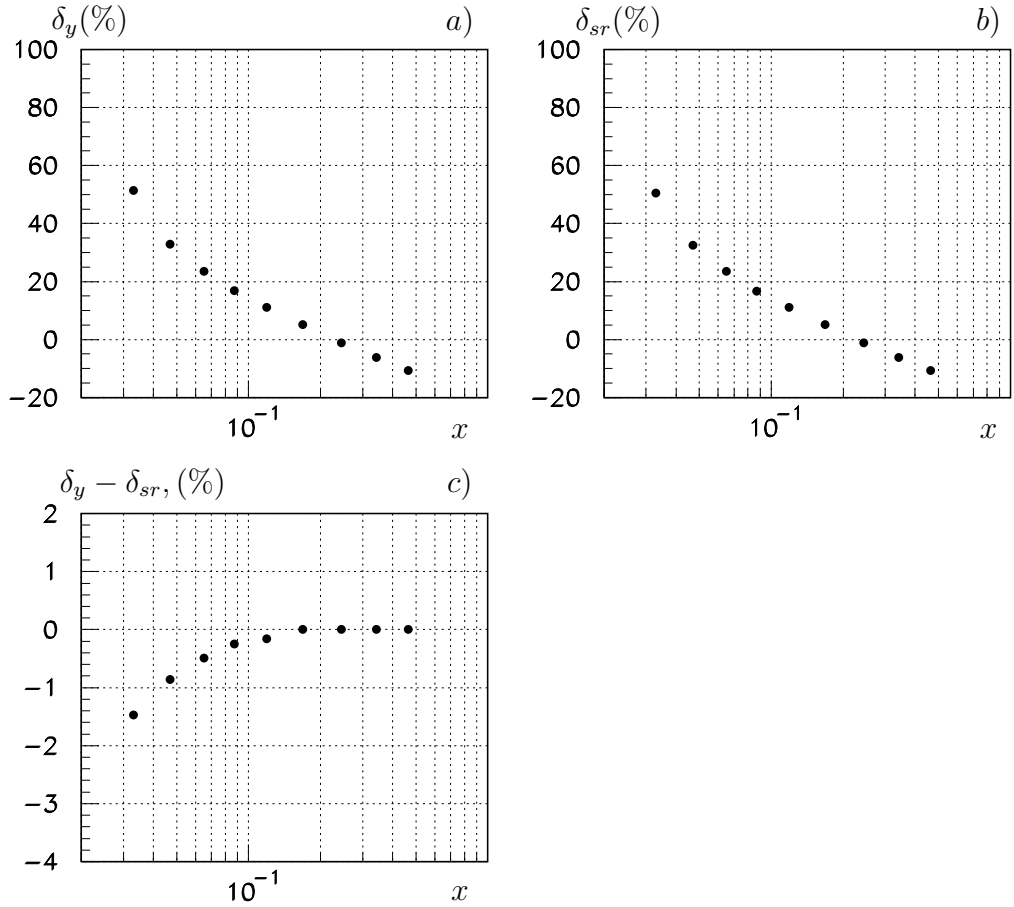}
}
\end{picture}
\vspace{7.5cm}
\caption{Relative correction to cross section $\delta $
calculated within the y-scaling hypothesis (2a),
within the sum
rules formalism (2b) and the difference (2c). HERMES kinematics
(see Table 1 of ref.\cite{Hermes}).
}
\label{figcs}
\end{figure}

\begin {equation}
\begin {array}{ll}
\displaystyle
f_A=1+2E_1/M_A\sin ^2(\bar\theta/2),
&\displaystyle
\eta_A=Q'^2_A/4M_A,
\\[0.3cm]\displaystyle
Q'^2_A=4E_1^2/f_A\sin ^2(\bar\theta/2),
&\displaystyle
\rho_A=1/(1+\eta_A),
\\[0.3cm]\displaystyle
\bar\theta=2\sin^{-1}\sqrt{{\tilde Q}^2M/(2E_1(ME_1-{\tilde Q}^2))},
&\displaystyle
q_a=\sqrt{Q^2_A\rho_A},
\end {array}
\label {0222}
\end {equation}
and
\begin {equation}
\begin {array}{cc}
\displaystyle
v_{TA}=
\textstyle \frac{1}{2}
\rho_A+\tan ^2(\bar\theta/2),
&\displaystyle
v'_{TA}=\tan (\bar\theta/2)\sqrt{\rho_A+\tan ^2(\bar\theta/2)}.
\end {array}
\label {0006}
\end {equation}
Correspondent quantities without index $A$ are obtained by
substitution
$M_A\longrightarrow M$.

\section{Numerical analysis for HERMES kinematics}

In the sections above we derived expressions for the contribution
of the QRT to the total correction of DIS on polarized $^3$He. All
quantities used, except for the suppression factors are well defined or
have been measured with good accuracy. The suppression factors thus
constitute the
main uncertainty in the calculation of the QRT. In this paper we
considered two different approaches to the calculation of the
suppression factors. The ${\tilde Q}^2$ dependence of suppression
factors calculated within the y-scaling hypothesis (fig.1a) and
within the sum
rules formalism (figs.1b - 1d) is presented in Fig.1. The predicted
behaviour for suppression factors is 1 for
big values of ${\tilde Q}^2$ and when
${\tilde Q}^2 \rightarrow 0$ suppression factors also goes to $0$.
The bend point correspond to the value ${\tilde
Q}^2=2k_f\approx0.1 GeV^2$, where $k_f$ is helium-3 Fermi
momentum.
As can be seen from Fig. 1 the difference in calculation comes
from the region below
the bend point, but this region is extremely important as due to
the behaviour of elastic formfactors with ${\tilde Q}^2$, the
region $M^2{\displaystyle x^2 \over \displaystyle 1-x} \lsim
{\tilde Q}^2 \lsim
2k_f$ gives
the biggest contribution to the calculation of the integral
(\ref{eq23q}).

\begin{figure}[t]
\hspace{2.5cm}
\begin{picture}(160,160)
\put(10,120){\makebox(0,0){$A_1$}}
\put(210,120){\makebox(0,0){$\Delta_y (\%)$}}
\put(10,-55){\makebox(0,0){$\Delta_{sr} (\%)$}}
\put(220,-55){\makebox(0,0){${\Delta_y - \Delta_{sr} \over A_1},
(\%)$}}
\put(150,-210){\makebox(0,0){$x$}}
\put(350,-210){\makebox(0,0){$x$}}
\put(150,-35){\makebox(0,0){$x$}}
\put(350,-35){\makebox(0,0){$x$}}
\put(150,120){\makebox(0,0){$a)$}}
\put(350,120){\makebox(0,0){$b)$}}
\put(150,-55){\makebox(0,0){$c)$}}
\put(350,-55){\makebox(0,0){$d)$}}
\put(0,-320){
\epsfxsize=13cm\epsfysize=18cm\epsffile{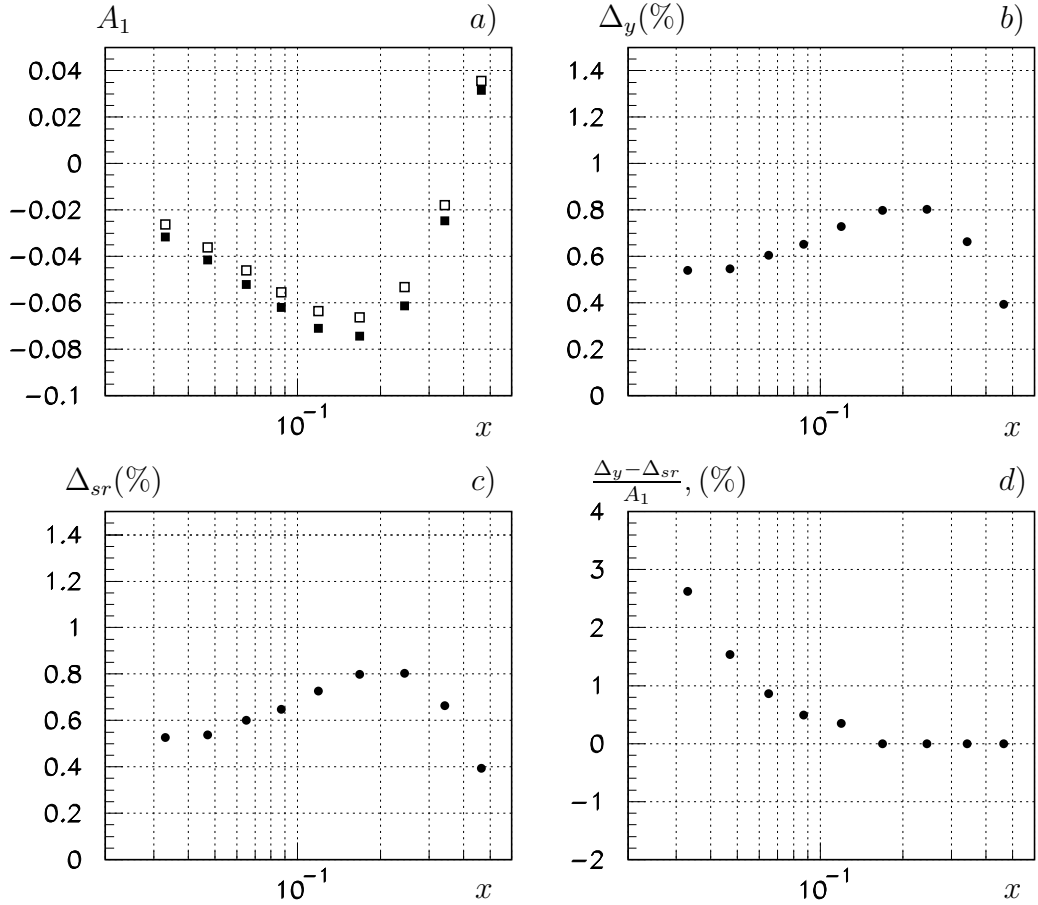}
}
\end{picture}
\vspace{7.5cm}
\caption{Born (open squares) and observed (full squares) $^3$He asymmetry (3a).
Relative correction to asymmetry $\Delta $
calculated within the y-scaling hypothesis (3b),
within the sum
rules formalism (3c) and the difference (3d) normalized over Born
asymmetry.
HERMES kinematics
(see Table 1 of ref.\cite{Hermes}).
}
\label{figas}
\end{figure}

To investigate the systematic error when measuring the observables
it is convenient to define quantities $\delta $  and $\Delta$:
\begin {equation}
\begin {array}{lr}
\displaystyle
\delta_y=\sigma^{obs}_y/\sigma_0 -1,
\displaystyle \;\;
\Delta_y=A^{obs}_{1y}-A_1^{born},
\\[0.2cm]\displaystyle
\delta_{sr}=\sigma^{obs}_{sr}/\sigma_0 -1,
\displaystyle \;\;
\Delta_{sr}=A^{obs}_{1sr}-A_1^{born}.
\end {array}
\label {0008}
\end {equation}
The indices $y$ and $sr$ correspond to observable quantities
calculated
within the y-scaling and sum rule approaches. The
spin $^3$He asymmetry is defined by the standard way
\begin{equation}
A_1=\frac{1}{D}{ \sigma^{\uparrow\downarrow} -
\sigma^{\uparrow\uparrow}\over
\sigma^{\uparrow\downarrow} + \sigma^{\uparrow\uparrow} } = 
{g_1(x,Q^2)\over F_1(x,Q^2)},
\end{equation}
and is roughly three times smaller the neutron spin asymmetry measured in
\cite{Hermes}. $D$ is the depolarization factor and $F_1$ and $g_1$ are
$^3$He structure functions.
    The results of numerical
calculations of the quantities (\ref{0008}) are presented in Figs.
2a-2b and
3b-3c. For spin-dependent structure function the model given in
Appendix of ref.\cite{gagu} is used. The kinematical
points (taken from Table 1 of ref.\cite{Hermes}) cover the
kinematical region of the HERMES experiment.
Large corrections occur mainly in the first bin due to the large values
of $y$ there.
Fig. 3a shows results for the asymmetry, with (observed) and without
(Born) the total RC.

It should be noted that the results
presented in Figs. 2c and 3d give an estimate of
the relative
systematic uncertainty on $^3$He spin dependent structure
functions due to the radiative tail from the quasielastic
peak (see also \cite{AKNA}). It is of the order of a few of percent in the
kinematical
region of modern polarization experiments. 

All numerical data were done for kinematical condition of HERMES, however
results for TJNAF are practically the same. The contribution of
quasielastic radiative tail is practically defined by $y$ and $x$
and
there is no essential dependence on incident electron energy. $y$ defines
the probability of the radiative subprocess, but $x$ fixes the low limit
of integration over $\tilde Q^2$ (or $\tau$).
We 
note also that due to integration main important region is relatively small
$Q^2$. For current experiments in DESY and TJNAF this region is
$Q^2<1$GeV$^2$.

\section{Discussion and Conclusion}

This paper is devoted to studying of radiative tail from quasielastic peak
in deep inelastic scattering on polarized and unpolarized $^3$He target.
Its contribution to total
RC is very important but the most insufficiently studied part of
standard procedure of radiative correction of experimental data. 
The main problem is that
contrary to the case of elastic formfactors and DIS structure functions
the $^3$He quasielastic response functions are not well studied yet. 

In our paper we tried to use some general properties of these functions
like y-scaling hypothesis, estimations of sum rules and simply a fact of
a presence of a peak at $Q^2=2M\nu$ in order to develop and  compare
different approaches
for  calculation of the correction. It is clear that the best formula is
exact expression (\ref{exact}), but it requires unknown information about
responses as functions of two variables $\nu$ and $Q^2$. So it is necessary
to look for some compromise between accuracy and uncertainties. 
We considered two possible approaches. First one is related to y-scaling 
hypothesis, second one gives an
expression in terms of sum rules for electron-nuclei scattering. The
results were analyzed and compared numerically. Difference between them
gives an estimation on systematical uncertainty due to quasielatic
radiative tail. We understand that consideration and comparison of only two 
approaches is not enough to make a conclusion about systematical
error. Unfortunately in literature there are no known to us
possibilities to construct any other approach which can provide 
explicit information for
some necessary quantities in $\vec {\rm e}^3\vec{\rm He}$ inclusive
quasielastic scattering. If such calculations or data will appear the
generalization of the developed approach is straightforward. 

For the current situation we recommend to use formulae (\ref{eq23q})
and (\ref{qf}) as basic to calculate radiative tail from quasielastic
peak. In our opinion the effect of the assumptions on these formulae is minimum. 
Furthermore these
assumptions can be controlled better and cannot lead to cruñial results
due to unknown ingredients. But further progress have to be connected
with exact formula (\ref{exact}). For that  new experimental
data or/and new reliable models have to appear.

\section*{Acknowledgments}

The work of IA was partially supported by the
U.S. Department of Energy under contract DE--AC05--84ER40150.



\Bibliography{99}
\bibitem {ST}
S.Stein et al.,
{Pys.Rev.} {\bf D12}(1975)1884
\bibitem {BA}
J.Bailey et al.,
{Nucl.Phys.} {\bf B151}(1979)367
\bibitem {MT}
L.Mo, Y.Tsai,
{ Rev.Mod.Phys.} {\bf 41}(1969)205;
\newline Y.Tsai,
SLAC-PUB-848, (1971).
\bibitem {ASh}
      I.V.Akushevich, N.M.Shumeiko, J.Phys., G {\bf20}(1994)513
\bibitem{Ak}
I.V.Akushevich, A.N.Ilyichev and N.M.Shumeiko,
J.Phys. G {\bf24}(1998)1995
\bibitem {HJM}
      P.Hoodbhoy, R.L.Jaffe and A.Manohar, Nucl.Phys.
 B{\bf 312}(1989)571
\bibitem {AT}
      I.Akushevich et al., Comp. Phys. Comm., {\bf 104}
(1997)201
\bibitem {LLS}
      W.Leidemann, E.Lipparini and S.Stringari, Phys.Rev.
{\bf C42}(1990)416
\bibitem {West}
G.B.West, Phys.Rep. {\bf18}(1975)263
\bibitem {Tho}
A.K.Thompson et al., Phys.Rev.Lett. {\bf68}(1992)2901
\bibitem {Wal}
T. de Forest and J.D.Walecka, Adv.Phys. {\bf15}(1966)1
\bibitem {Mon}
E.J.Moniz Phys.Rev. {\bf184}(1969)1154
\bibitem {OrTr}
      G.Orlandini and M.Traini, Rep.Prog.Phys. {\bf54}(1991)257
\bibitem {Shia}
R.Schiavilla, D.S.Lewart, V.R.Pandharipande,
S.C.Pieper, R.B.Wiringa, and S.Fantoni
Nucl. Phys. {\bf A473}(1987)267
\bibitem {Hermes}
K.Ackerstaff et al.,
Phys. Lett. B{\bf 404}(1997)383
\bibitem{gagu}
N.~Gagunashvili et al.,
Nucl.\ Instrum.\ Meth.\  {\bf A412} (1998) 146.
\bibitem{AKNA}
I.~Akushevich and A.~Nagaitsev,
J.\ Phys. {\bf G24} (1998) 2235
\endbib
\end{document}